\documentclass[a4paper,12pt]{article}

\usepackage{graphics}

\begin{document}

\title{
\begin{flushright}
{\normalsize IRB-TH-6/03}
\end{flushright}
\vspace{2 cm}
\bf Generalized phantom energy}

\author{ H. \v Stefan\v ci\'c\thanks{shrvoje@thphys.irb.hr}
}

\vspace{3 cm}
\date{
\centering
Theoretical Physics Division, Rudjer Bo\v{s}kovi\'{c} Institute, \\
   P.O.Box 180, HR-10002 Zagreb, Croatia}

%\institute{
%  Theoretical Physics Division, Ru\dj er Bo\v{s}kovi\'{c} Institute,
%   P. O. Box 1016, HR-10001 Zagreb, Croatia}

\maketitle

\abstract{We examine cosmological models with generalized phantom energy (GPE).
Generalized phantom energy satisfies  the supernegative equation of state, 
but its evolution with
the scale factor is generally independent, i.e. not determined by its equation
of state. The requirement of general covariance makes the gravitational
constant time-dependent. It is found that a large class of distinct GPE
models with different evolution of generalized phantom energy density and
gravitational constant, but the same equation of state of GPE have the same
evolution of the scale factor of the universe in the distant future.
The time dependence of the equation of state parameter determines
whether the universe will end in a de Sitter-like phase or diverge in finite
time with the accompanying ``Big Rip" effect on the bound structures.}

\vspace{2cm}

Results of recent cosmological observations, such as distant supernovae of type
Ia (SNIa) \cite{SN} and cosmic microwave background radiation (CMBR)
\cite{CMBR}, have dramatically altered our perception of the dynamics and 
composition of the universe and reshaped the landscape of standard cosmology
\cite{Rev}. The universe seems to be in the 
phase of 
accelerated expansion, which started at a relatively small redshift, $z \sim 1$.
This acceleration is attributed to a new form of matter,
usually referred to as {\em dark energy}, the nature of which is still not
definitely established. Observations indicate that the energy density of the
universe is very close to its critical density where dark energy presently
accounts for approximately $2/3$ of the total energy density, while the
remaining $1/3$ comes predominantly from {\em dark matter}, another
unidentified component of the universe.  
The most prominent and studied candidates 
for the title of dark energy are the cosmological constant \cite{wein,peeb,pad}
(together with its dynamical variants, such as renormalization group running
cosmological constant \cite{mi,sola,bonanno}), quintessence \cite{Q} and the 
Chaplygin gas \cite{CG}. 
%The cosmological constant 
%comprises various
%contributions of vacuum energy density. Although it is favoured
%because of its conceptual simplicity and traditional reasons, 
%the cosmological constant is burdened by 
%the necessity of the huge fine-tuning between the size of the particle physics 
%contributions and the size of the measured dark energy density. 
%Another problem is why
%the energy density of the cosmological constant is 
%comparable with the matter energy
%density at the present epoch, which would, given the distinct scaling of
%cosmological constant and matter with the scale factor, require another 
%fine-tuning in the distant past. Some of these problems could be alleviated in
%the formalism of dynamic cosmological constant, such as the renormalization
%group running cosmological constant \cite{mi, sola, bonanno}. In quintessence
%models dark energy is described by the scalar field evolving in a potential.
%The nature of the quintessence field and the exact form of the 
%potential are still
%not unambiguously determined. The Chaplygin gas is the model based on a fluid with
%an unorthodox equation of state. One of its most prominent features is the
%scaling of its energy density with the scale factor which makes it a candidate
%for the unification of dark energy and dark matter.  
%A multitude of other, related and unrelated,
%dark energy models exists, reflecting the complex and multifaceted nature of
%dark energy itself.           

The majority of dark energy models share a common constraint on their
equation of state ($p_{d}$ and $\rho_{d}$ represent pressure and energy density
of dark energy, respectively)
\begin{equation}
\label{eq:eos}
p_{d} = w \rho_{d} \, ,
\end{equation}
where $w \ge -1$. Such a constraint is, however, not justified by the unbiased
fits to the data of cosmological observations. Moreover, the allowed interval 
for the parameter of the equation of state extends significantly into the region 
with $w < -1$. The use of observational data on CMBR, large scale
structure (LSS), SNIa and Hubble parameter measurements from the Hubble Space
Telescope (HST) under the assumption of the redshift independent 
parameter $w$ give the restriction $-1.38 < w < -0.82$ at the 95\% confidence
level \cite{odman}. 
Therefore, a possible supernegative equation of state of dark energy 
deserves due attention.

A new type of dark energy with the equation of state characterized by 
$w < -1$ was proposed in \cite{cald} and named {\em phantom energy}. 
Phantom energy is considered to be separate from other components of the
universe and its energy-momentum tensor is conserved separately. In such a
setting, the equation of state of dark energy determines its evolution with
the scale factor $a$. The
supernegative nature of the equation of state of the phantom energy leads to
the growing energy density of phantom energy $\rho_{d} \sim a^{-3(1+w)}$,
for a constant parameter $w$.
The cosmological dynamics of the universe with such a phantom energy component
possesses many interesting features \cite{cald2}. 
The growth of the energy density of phantom energy drives the scale factor
of the universe to infinity in finite time. The increasing negative pressure of
phantom energy leads to the unbounding of all bound structures in the universe.
This dramatic and picturesque scenario of the cosmic doomsday was appropriately
named ``Big Rip". The formulation of microscopic models for 
phantom energy \cite{tech}
relies on the machinery developed in quintessence models, namely the evolution
of the scalar field in a suitably chosen potential. 
However, the description of phantom energy may require an 
introduction of some nonstandard alterations, e.g. the negative
kinetic term of the scalar field. Detailed considerations of the Lagrangians
describing phantom energy show that in some cases the universe
with phantom energy ends in a ``Big Rip", while in others it asymptotically
approaches the de Sitter expansion.

In this paper, we consider models with generalized phantom energy (GPE). 
First, we set up a more general model of the evolution of the universe with
phantom energy. 
%The main difference compared to the standard approach is
%that we allow the variation of the gravitational coupling $G$ with time. 
%Models with the time dependent $G$ were extensively studied in the framework of
%the time dependent cosmological term $\Lambda(t)$ \cite{Gvar}. 
We
assume that there are two components of the universe: the dark energy component
(which will have the phantom energy characteristics), and the ``ordinary" matter
component with the respective energy densities $\rho_{d}$ and $\rho_{m}$. The
``ordinary" matter is taken to satisfy the equation of state
\begin{equation}
\label{eq:eosmatter}
p_{m}=\gamma \rho_{m} \, ,
\end{equation}
where $\gamma \ge 0$.
Furthermore, we assume that the energy-momentum tensor of the ``ordinary" 
matter is conserved
\begin{equation}
\label{eq:Tmatter}
T^{\mu \nu}_{m ; \nu} = 0   \, .
\end{equation} 
The equation given above ensures that the parameter of the equation of state
governs the evolution of the ``ordinary" matter energy density, i.e.
\begin{equation}
\label{eq:rhomatter}
\rho_{m} = \rho_{m,0} \left( \frac{a}{a_{0}} \right)^{-3(1+\gamma)}   \, .
\end{equation}
Dark energy has the equation of state
\begin{equation}
\label{eq:eosdark}
p_{d}=w \rho_{d} \, ,
\end{equation}                  
where $w$ generally depends on time explicitly or implicitly, via explicit
dependence on some other time-dependent quantity, such as the scale factor $a$. 
In the case of dark
energy, we allow the possibility of non-conservation of the energy-momentum 
tensor, i.e.
\begin{equation}
\label{eq:Tdark}
T^{\mu \nu}_{d ; \nu} \neq 0   \, .
\end{equation}  
Thus, the evolution of the dark energy density is not determined by the parameter
from its equation of state.

With the properties of the components of the universe defined, we can specify
the laws of its evolution. We start from the Einstein equation 
\begin{equation}
\label{eq:Einsteinlaw}
G^{\mu \nu} = -8 \pi G T^{\mu \nu} \, ,
\end{equation}    
where $G^{\mu \nu}$ is the Einstein tensor and $T^{\mu \nu} = T^{\mu \nu}_{d}+
T^{\mu \nu}_{m}$ is the total energy-momentum tensor. The reconciliation of the
requirement of the general covariance of (\ref{eq:Einsteinlaw}) and the
non-conservation relation (\ref{eq:Tdark}) is possible with the promotion of
gravitational constant $G$ into a space-time dependent quantity. This change can
be interpreted as a modification of the dynamics of General Relativity. This
additional dynamics is effectively described by the introduction of space-time
dependence of $G$. We consider the models where $G$ is a function of time only,
$G = G(t)$.
Models with the time-dependent $G$ were extensively studied in the framework of
the time-dependent cosmological term $\Lambda(t)$ \cite{Gvar}.
The covariant derivative of (\ref{eq:Einsteinlaw}) then  implies
\begin{equation}
\label{eq:gencov}
(G(t) T^{\mu \nu})_{;\nu} = 0 \, .
\end{equation}
This equation can be rewritten in the form
\begin{equation}
\label{eq:gencons}
d(G(\rho_{m}+\rho_{d})a^{3}) = - G(p_{m}+p_{d})d a^{3} \, .
\end{equation}
Combining the evolution laws (\ref{eq:rhomatter}) and (\ref{eq:gencons}) 
and introducing $w \equiv -1 + \kappa$ (where $\kappa$ describes the deviation
from the parameter of the equation of state inherent to the cosmological
constant) we arrive at
\begin{equation}
\label{eq:Gevolution}
\dot{G}(\rho_{m}+\rho_{d}) + G \dot{\rho_{d}} + 3 \kappa H G \rho_{d} = 0 \, .
\end{equation}
Here $H = \dot{a}/a$ is the Hubble parameter, while dots
denote time derivatives. Equation
(\ref{eq:Gevolution}) clearly shows the generality of the model. In the case of
the constant $G$, we recover the standard equation of conservation of 
$T^{\mu \nu}_{d}$. Equation (\ref{eq:Gevolution}) shows that the time evolution
of $G$ is the result of two competing effects. Namely, for dark energy with
growing energy density, the second term in (\ref{eq:Gevolution}) causes the
decrease of $G$, while for negative $\kappa$, the third term in 
(\ref{eq:Gevolution}) increases $G$ with time.

Finally, Friedmann equations for the evolution of the scale factor complete the 
set of evolution equations (\ref{eq:rhomatter}) and (\ref{eq:Gevolution})
\begin{equation}
\label{eq:Friedmann}
\left(\frac{\dot{a}}{a} \right)^{2} + \frac{k}{a^{2}} = \frac{8 \pi}{3}
G (\rho_{m}+\rho_{d})\, ,
\end{equation}
\begin{equation}
\label{eq:Friedmannacc}
\frac{\ddot{a}}{a} = - \frac{4 \pi}{3} G (\rho_{m}+\rho_{d} + 3p_{m} + 3p_{d})
\, .
\end{equation}

The set of equations (\ref{eq:rhomatter}), 
(\ref{eq:Gevolution}) and (\ref{eq:Friedmann}) 
reveals that we have essentially two independent equations
for three dynamical quantities $G$, $\rho_{d}$ and $a$ (assuming that $\kappa$
is the function of these quantities and time). 
Without a more specific identification of
the dynamics of $G$ or $\rho_{d}$, it is not possible to solve the aforementioned
set of equations. However, as we show below, with mild assumptions about the 
evolution of dark energy with the scale factor, it is possible to obtain information
on the future evolution of the universe for general $G$ and $\rho_{d}$ 
satisfying the equations given above.

Next, we introduce the concept of generalized phantom energy (GPE). 
Generalized phantom energy is the form of dark energy satisfying the equation of
state (\ref{eq:eos}) with the non-conserved 
energy-momentum tensor (\ref{eq:Tdark})
and the following two properties:
\begin{enumerate}
%\roman{enumi}
\item[(a)] {\em GPE energy density is a non-decreasing function of the 
scale factor,}
\item[(b)] {\em GPE equation of state satisfies $\kappa \le 0$.} 
\end{enumerate}

We further examine the future evolution of the universe. 
 In the sufficiently distant future we have
$\rho_{m} \ll \rho_{d}$ and $\rho_{m}$ can be neglected in the evolution
equations. Equations (\ref{eq:Gevolution}) and (\ref{eq:Friedmann}) thus become
\begin{equation}
\label{eq:norhomF}
\left(\frac{\dot{a}}{a} \right)^{2} + \frac{k}{a^{2}}  =  \frac{8 \pi}{3}
G \rho_{d} 
\end{equation}
and
\begin{equation}
\label{eq:norhomG}
\frac{d}{dt}(G \rho_{d}) + 3 \kappa H G \rho_{d} = 0 \, . 
\end{equation}   

Furthermore, from equation (\ref{eq:norhomG}), we obtain
\begin{equation}
\label{eq:Grhod}
\frac{d(G \rho_{d})}{G \rho_{d}} = - 3 \kappa \frac{da}{a} \, . 
\end{equation} 
As the condition $-\kappa \ge 0$ is satisfied by assumption (b), we obtain
\begin{equation}
\label{eq:growGrhod}
G \rho_{d} \ge (G \rho_{d})_{0} \, . 
\end{equation}
Therefore, as $G \rho_{d}$ is a growing function in an expanding universe, 
for large $a$ we can
disregard the term $k/a^{2}$ in equation (\ref{eq:norhomF}). For the flat
universe, this approximation is exact, while for the closed or the open
universe,
this approximation is applicable in the sufficienly distant future.

Finally, we end up with the following two equations for the dynamics of the
universe in the distant future:
\begin{equation}
\label{eq:distdyn1}
H^{2}  =  \frac{8 \pi}{3} (G \rho_{d}) \, , 
\end{equation}
\begin{equation}
\label{eq:distdyn2}
\frac{d}{dt}(G \rho_{d}) + 3 \kappa H (G \rho_{d})  =  0
\, . 
\end{equation}
By combining equations (\ref{eq:distdyn1}) and (\ref{eq:distdyn2}), we obtain an
equation for the evolution of the Hubble parameter $H$ with time
\begin{equation}
\label{eq:evolH}
\frac{dH}{dt} + \frac{3}{2} \kappa H^{2}   =  0 \, , 
\end{equation}
with the solution
\begin{equation}
\label{eq:soluH}
H(t) = \frac{H(t_{0})}{1+\frac{3}{2} H(t_{0})
\int_{t_{0}}^{t} \kappa(t') dt'} \, . 
\end{equation}
Once we have found the expression for the evolution of the Hubble parameter,
 it is easy to obtain an expression for the evolution of the scale factor $a$
\begin{equation}
\label{eq:solua}
a(t) = a(t_{0}) exp \left( \int_{t_{0}}^{t} dt' \frac{H(t_{0})}{1+\frac{3}{2} 
H(t_{0}) \int_{t_{0}}^{t'}  \kappa(t'') dt''} \right) \, . 
\end{equation}

General solutions (\ref{eq:soluH}) and (\ref{eq:solua}) exhibit some interesting
features. The evolution of the universe in the sufficiently distant future is
governed only by the parameter of the equation of state of dark energy. The
precise form of the growth of $\rho_{d}$ with the scale factor $a$ is  
irrelevant in this limit. This implies that the entire class of models with 
different functional forms of $\rho_{d}$ and $G$, 
obeying the same equation of state, show the same
behaviour in the sufficiently distant future.  Therefore, we can divide
all GPE models with the characteristics specified above into
classes with the same equation of state.

An important question regarding the fate of the universe is whether, for 
a particular class
of generalized phantom energy models, $a$ and $H$ diverge in finite time or
reach infinite values only in infinite time. For the Hubble
parameter $H$, the answer is straightforward. There will be no divergence of $H$
in finite time if the denominator of the expression on the right-hand side of 
(\ref{eq:soluH}) remains positive for all times. This leads to the condition
\begin{equation}
\label{eq:cond} 
\int_{t_{0}}^{\infty}  (-\kappa(t')) dt' <  \frac{2}{3 H(t_{0})} \, . 
\end{equation}
As in this case there is no singularity in $H(t)$ in finite time, the scale
factor $a(t)$ also does not diverge in finite time.
In order to have the convergence of the integral 
$\int_{t_{0}}^{\infty}  (-\kappa(t')) dt'$ required in (\ref{eq:cond}), 
the function $\kappa(t)$ has to tend to zero at asymptotically large times.
Therefore, for generalized phantom matter which exhibits no divergence of $H$
or $a$ in finite time, the parameter of the equation of state approaches $-1$,
i.e. generalized phantom energy approaches the time-dependent cosmological
term.

In the case when the condition (\ref{eq:cond}) is not satisfied, the Hubble
parameter $H$ diverges in finite time $t$. From Friedmann equations we have
\begin{equation}
\label{eq:inf1} 
\dot{a} = H a \, , 
\end{equation} 
\begin{equation}
\label{eq:inf2} 
\ddot{a} = \left(1 - \frac{3}{2}\kappa \right) H^{2} a \, . 
\end{equation}
These expressions indicate that, when $H$ diverges in finite time $t$, 
both $\dot{a}$ and $\ddot{a}$ diverge as well, so the scale factor $a$ cannot
remain finite, but diverges in finite time $t$ as well. 

From the general expessions (\ref{eq:soluH}) and (\ref{eq:solua}), we can obtain
evolution laws for the conceptually simple, but important case \cite{cald}
\begin{equation}
\label{eq:const} 
\kappa(t) = - \kappa_{0}  \, . 
\end{equation}
With such a choice for the parameter of the equation of state of generalized
phantom energy, we have the following evolution laws:
\begin{equation}
\label{eq:Hconst}
H(t) = \frac{H(t_{0})}{1-\frac{3}{2} H(t_{0}) \kappa_{0} (t-t_{0})} \, , 
\end{equation}
\begin{equation}
\label{eq:aconst}
a(t) = a(t_{0}) \left( 1 - \frac{3}{2} H(t_{0}) \kappa_{0} (t - t_{0})
\right)^{-\frac{2}{3 \kappa_{0}}} \, .
\end{equation}
These solutions clearly show the onset of the divergence in $H$ and $a$.
The universe with generalized phantom energy with the constant parameter of the
equation of state evolves to infinity in finite time. 

Comparison with the case of the ``standard" phantom energy
\cite{cald, cald2} shows that, for the same parameter of the equation of state
$\kappa(t)$, the scale factor follows the same evolution law. Given the fact
that the parameter of the equation of state does not determine the scaling with
$a$, and that $G$ is variable in the framework of generalized phantom
energy,
it is by no means obvious that coincidence of this sort should exist. However,
from the equation (\ref{eq:Gevolution}),
 we readily see that for the case of constant $G$, we recover
the equation of evolution for the ``standard" phantom energy. 
As far as the evolution in the sufficiently distant future
is concerned, the ``standard" phantom
energy model is just one instance of the class of generalized 
phantom energy models with the same function $\kappa(t)$.     

Given the same evolution properties of the broad class of GPE
models with the same $\kappa(t)$, it is natural to look at the destiny of
bound structures, another peculiarity of phantom energy models \cite{cald2}.
The relevant quantity with respect to the stability of the bound structures is
the analogue of the gravitational potential proportional to the quantity 
$G(\rho_{d}+3p_{d}) = (-2 + 3\kappa) G\rho_{d}$.  Equation (\ref{eq:distdyn1}) 
shows that $G\rho_{d} \sim H^{2}$ and $G\rho_{d}$ grows with
time. If the condition (\ref{eq:cond}) is not satisfied, $H$ and
$G\rho_{d}$ diverge in finite time. Furthermore, as $\rho_{d}$ grows with the
scale factor, $G \rho_{d}$ certainly increases compared to $G$. For
gravitationally bound systems, the GPE contribution of the order 
$ \sim G(\rho_{d}+3p_{d}) R^{3}$ (where $R$ denotes the characteristic 
spatial scale of the bound system) overwhelms the ``mass" contribution $\sim G
M$ ($M$ denotes the mass of the bound system). Gravitationally bound systems
fall apart in finite time. For the systems bound by electromagnetic or strong
forces, mere growth of $G\rho_{d}$ ensures their unbounding at some finite time
before the time at which scale factor goes to infinity.
Consequently, all bound structures are
unbound in finite times. The scenario of the ``Big Rip" is present in 
generalized phantom energy models as well.

Finally, let us make some comments on fundamental aspects of the GPE model.
As the gravitational constant $G(t)$ is time-dependent, the description of
the gravitational sector in the GPE model represents a declination from the
Einsteinian gravity. One important aspect is whether the scale factor $a$ really
describes the growth of length scales. One can raise two arguments in favour of
the standard interpretation of the scale factor $a$. The first is that no
intervention in the geometrical structure or interpretation of the left-hand
side of equation (\ref{eq:Einsteinlaw}) has been made. The other, more physical one, is that
the density of ``nonrelativistic" matter scales as $\rho_{m} \sim a^{-3}$ in
our GPE model, equation (\ref{eq:rhomatter}) with $\gamma = 0$. Given that no 
interaction (production or annihilation) of the
``ordinary" matter component with other components is assumed, this fact
establishes $a$ as a natural measure of the growth of length scales. 

In some theories with the time-dependent effective gravitational constant, such
as scalar-tensor or nonminimally coupled scalar field theories, one can
construct many mathematically equivalent theories using conformal
transformations. It turns out that all these theories are not physically
equivalent, i.e. some formulations are more physically viable than others
(the Einstein frame formulation is more viable than the Jordan frame formulation)
\cite{Faraoni}. Generally, it might
be of interest to consider conformally related models of GPE obtained by the
transformation of the type $\tilde{g}_{\mu\nu} = f(G(t)) g_{\mu\nu}$, where $f$
is a suitably chosen function.
However, the time variation of $G(t)$ in our model can be very general and
includes possibilities to which requirements on the choice of the 
conformal frame do
not necessarily apply. Some examples of such a variation are the renormalization
group running of $G$ \cite{mi, sola, bonanno} 
or the time variation of $G$ emanating from extra dimensions \cite{extra}.       

In conclusion, in this letter we have considered cosmological models with the 
time-dependent gravitational constant $G$ and dark energy with the supernegative
equation of state (phantom energy). Phantom energy is generalized in the
sense that its equation of state does not determine its evolution with
the scale factor $a$, i.e. GPE density becomes an independent function of the scale factor.
The requirement of general covariance in this setting imposes conditions on
the gravitational constant $G$ which acquires time dependence. Investigation of 
future dynamics of the generalized phantom energy models with growing
generalized phantom energy density and the parameter of the equation of state
less than $-1$ exhibits some general properties. A large class of models with
different evolutions of $\rho_{d}$ and $G$, but the same equation of state of 
GPE, have the common law of the evolution of the scale
factor $a$ in the sufficiently distant future. 
The time dependence of the GPE parameter
of the equation of state determines whether the
universe evolves infinitely in a de Sitter regime or diverges in finite time.
One would expect that bounds on the variation of $G$ in
the past epochs of the evolution of the universe would produce the most stringent
constraints on the parameters of the GPE model. Therefore,
it is important to point out that our main results qualitatively do not depend
on the size of the parameter $|\kappa|$ or on the intensity of growth of $\rho_{d}$
(of course, within classes of these parameters that satisfy or do not satisfy
the condition (\ref{eq:cond})). For smaller parameter values and 
slowlier varying
functions $\rho_{d}$ and $G$, the onset of the general evolution (dependent only
on $\kappa$) will come later. For instance, for constant and negative $\kappa$,
but very small $|\kappa|$, the entire class of GPE models leads to the ``Big Rip"
event, but at very late times. 

Clearly, the present accelerating phase of the
evolution of the universe carries the seed of the possibly very dramatic future
of our cosmos. Therefore, more precise observations of the past
variation of $\rho_{d}$ and $G$ with time (redshift) will 
be able to unravel the fate of the universe.      

{\bf Acknowledgements.} The author would like to thank N. Bili\'{c}, B.
Guberina and R. Horvat for useful comments on the manuscript. This work was
supported by the Ministry of Science and Technology of the Republic of Croatia 
under the contract No. 0098002.

\end{document}